\documentclass{article}

\usepackage[nonatbib,final]{neurips_2020}

\usepackage[utf8]{inputenc} 
\usepackage[T1]{fontenc}    
\usepackage{hyperref}       
\usepackage{url}            
\usepackage{booktabs}       
\usepackage{amsfonts}       
\usepackage{nicefrac}       
\usepackage{microtype}      

\title{An Ethical Highlighter for People-Centric\\Dataset Creation}

\author{%
  Margot Hanley$^{1,2}$ \qquad Apoorv Khandelwal$^{2}$ \qquad Hadar Averbuch-Elor$^{1,2}$ \\ \textbf{Noah Snavely$^{1,2}$ \qquad Helen Nissenbaum$^{1,2}$}
  \vspace{3px} \\
  $^1$Cornell Tech \hspace{10pt} $^2$Cornell University \\
  \texttt{\{mh2446,ak2254,hadarelor,snavely,hn288\}@cornell.edu}
}

\begin{document}

\maketitle

\begin{abstract}
Important ethical concerns arising from computer vision datasets of people have been receiving significant attention, and a number of datasets have been withdrawn as a result.
To meet the academic need for people-centric datasets, we propose an analytical framework to guide ethical evaluation of existing datasets and to serve future dataset creators in avoiding missteps. 
Our work is informed by a review and analysis of prior works and \emph{highlights} where such ethical challenges arise.

\end{abstract}

\section{Introduction}
\label{sec:introduction}
In recent years, face datasets from the computer vision community have seen significant criticism, and many have subsequently been withdrawn by their creators. 
Why? In prioritizing properties like size and downstream task utility, other principles and factors have often taken a backseat, such as data curation practices, privacy violations, offensive labels, fair representation, and undesirable uses. Although datasets are often created with the positive aim of furthering scientific research, the ethical challenges prompting these takedowns reveal problematic (and often unintended) dataset properties and consequences. These issues are not limited to face datasets---e.g. the ImageNet dataset has been charged with offensive labels (inherited from WordNet~\cite{crawford2019excavating,yang2020towards}) and has faced other issues regarding its ``Person'' category~\cite{deng2009imagenet}.

Although the dataset takedowns have exposed complex ethical challenges, the need for publicly available, large-scale datasets remains critical to open academic research. These datasets may serve common and positive needs, such as synthesizing people for privacy enhancing applications, image-to-text descriptions in support of accessibility, and computer vision--driven methods for revealing demographic biases in media~\cite{google2017thewomen,hong2020analyzing}. Hence, the question remains---how can the community promote pro-social academic research through the creation and evaluation of critical data resources that can stand the test of time, given the serious ethical challenges faced by prior datasets?

Our work addresses this question by proposing an analytic framework for evaluating image datasets of faces, 
people, and other scenes featuring people---hereafter, \textit{people-centric datasets}---and by providing guidance to dataset creators seeking to address ethical standards alongside important technical considerations. The framework, which we refer to as an ``ethical highlighter'', comprises four components, each encapsulating a distinctive aspect of dataset construction where ethical challenges can arise: creation, composition, distribution, and purpose. 

Our work shares common ground with prior efforts such as Datasheets for Datasets~\cite{gebru2018datasheets}, a significant milestone in drawing attention to transparency and accountability in dataset construction. Building on these efforts, our framework crystallizes distinct aspects of dataset development and draws explicit threads between these aspects and ethical issues, thereby offering computer vision researchers the means for not only committing to ethical standards in principle, but also meeting them in practice.

\section{Framework}
\label{sec:framework}
\noindent \textbf{Methodology.}
After reviewing many people-centric datasets, we selected 14 as an initial sample set---these datasets have either been widely-used or drawn prominent public criticisms on ethical grounds. While all had been publicly available at some point, 11 have since been taken down and only three remain available. Through an inductive analysis of academic manuscripts and popular press~\cite{megapixels}, we produced a typology of critiques. While this was not an exhaustive review, our team collected this work to a point of conceptual saturation~\cite{glaser1968discovery}.

From our inductive analysis, we extracted two key dimensions around which the implicit and explicit themes we encountered could be organized. We identified \textbf{components of dataset development} as Creation, Composition, Distribution, and Purpose. And, we identified a non-exhaustive list of \textbf{types of ethical issues} which emerged through our inductive analysis: fairness, privacy, subject autonomy, safety/security, property rights, representational harm, offense, transparency/explainability, and accountability/responsibility. In some instances, we framed these concerns according to  traditional ethical concepts, which are commonly cited in broader discussions of algorithmic and AI systems. We sought to understand how the ethical issues were dispersed across and manifested in our framework's components. Due to space limitations in this short article, we briefly describe the components and illustrate each with one ethical issue. However, a fuller version in the future will provide a more comprehensive analysis, as well as critical reflections that can be used to guide dataset creators.

\subsection{Creation}
\label{sec:creation}
\emph{Creation} encompasses the activities involved in producing a dataset, including sourcing, assembling, and cleaning data, as well as assigning labels. Typically, image datasets comprise images and their associated labels. These labels could be names of the people pictured, attributes (e.g., age), or potentially any other descriptions. More generally, we can think of labels as data in a textual modality associated with data in a visual modality (for now we exclude annotations like segmentation or depth maps). Our inductive analysis of this \emph{creation} component revealed ethical concerns over violations of privacy and property rights, offensive data content, and subject autonomy. 

\medskip
\noindent \textbf{Privacy.}
An early decision that dataset creators make is where to source the data. Broadly, this choice can be characterized as capturing images of subjects directly or sourcing images secondhand (e.g. scraping them from social media sites or search engines). 

In order to amass large numbers of images of people, some dataset creators have previously captured photos and videos without subjects' knowledge or consent, two important aspects of privacy. For example, the creators of Duke MTMC~\cite{ristani2016performance} and Brainwash~\cite{stewart2016end} were criticized for assembling datasets without notice or consent. Duke MTMC, created in 2014, comprised live footage of students on campus. In the same year, the Brainwash dataset used a webcam to capture images of customers in the San Francisco \textit{Brainwash} cafe. In response to these criticisms, both datasets were ultimately taken down by their authors, respectively in May and June of 2019. The authors of Duke MTMC acknowledged that they had deviated from IRB-approved protocols by filming outdoors and releasing data without protections~\cite{duke_website}.

Even if dataset creators aim to capture faces in natural environments, they must grapple with the question of consent and determine methods to mitigate other privacy concerns. The creators of UnConstrained College Students (another dataset created on a college campus) suggested that some of its value was from featuring subjects that were ``photographed [at long-range] without [students'] knowledge'', rather than being ``posed''~\cite{opensetface}. Images collected ``in the wild'' are considered to be particularly useful across a variety of naturalistic application domains. From the perspective of these creators, a lack of subject awareness (leading to a lack of consent) has been a feature, rather than a bug. It is also important to keep in mind that even when subjects give their consent to be included, they do not necessarily consent to all possible uses of a dataset---creators should be careful to specify the scope of the dataset's purposes to subjects.

Another common method for gathering raw image data is to find and download images of people from the internet. For this method, it seems to be common practice to assume that ``public persons'' cede the right to privacy. For example, the creators of MS-Celeb-1M~\cite{guo2016msceleb} started by assembling a list of 1M ostensible celebrities, selecting a subset of 100K identities and scraping their images from search engines. However, the dataset creators' definition of ``celebrity'' was very broad---the original list was not limited to individuals who consider themselves as public figures, but was rather one million notable people on the internet, including a number of private persons. Consequently, when the dataset was released, it emerged that many faces were not those of celebrities, but instead of non-public persons (including vocal privacy advocates and journalists) who had not consented to having their faces included in the dataset.

\subsection{Composition}
\label{sec:composition}
By \emph{composition}, we refer to properties of the dataset, spanning content (e.g., data units or elements comprising the dataset) such as visual images and text-based labels, mappings among elements expressed in different modalities (e.g., labels to images), and higher-order, macro attributes of the dataset such as demographic representativeness. Our analysis revealed that the composition of a dataset may be a source of ethical harms through, e.g., bias and unfairness, representational harm, and offensiveness.

\medskip
\noindent \textbf{Offense.}
Offensive associations can be latent in popular machine learning datasets. A notable example where such associations were made visible was through a web-based demonstration called ``ImageNet Roulette'', created by researcher-artists Kate Crawford and Trevor Paglan via training on the full ImageNet dataset~\cite{deng2009imagenet}. By allowing users to upload images of themselves and publishing the resulting classifications, the project exposed shocking labels attributed to ImageNet imagery (which contains many people categories, unlike the subset used in the well-known ILSVRC challenge) to a general audience. For instance, labels included ``rape suspect'', ``pipe smoker'', ``alcoholic'', and ``bitch''. People of color could potentially be labeled with racial slurs. Just one week following the launch of this project, the ImageNet team took down the ``Person'' category for maintenance.

Sources of offensive associations have been traced to labels used to generate ImageNet---namely, a database of words and semantic relations called WordNet~\cite{fellbaum2012wordnet}. In their analysis of the ImageNet subtree, Yang et al.\ (including ImageNet team members) found that of the 2,832 people categories within the subtree, 1,593 were potentially offensive categories~\cite{yang2020towards}. Databases like Wordnet are often used to build datasets and are adopted across the industry. Other datasets built on WordNet, such as 80 Million Tiny Images~\cite{torralba200880}, have similarly inherited offensive associations from WordNet. However, in this case, the $32 \times 32$ images featured content that would be too small to manually perceive and audit~\cite{prabhu2020large}. Indeed, datasets like ImageNet have included images that are offensive or portray certain sub-populations in perjorative ways. In order to prevent such offensive images and associations, there is a need for more thorough auditing of both labels and imagery~\cite{kyriakou2019fairness,barlas2019social}.

\subsection{Distribution}
\label{sec:distribution}
\emph{Distribution} is concerned with how creators make a dataset available, as well as that dataset''s terms of use and disclaimers. Our analysis revealed that the distribution of a dataset presents a source of ethical harms when it impedes accountability and violates subject autonomy.

\medskip
\noindent \textbf{Accountability (responsibility).}
Even if creators have the best intentions for their dataset, they must prepare for the possibility that users will not use it for its intended purpose. As part of the provisions of access, many datasets request that users use data only for non-commercial research purposes, but are unable to enforce this usage once third parties obtain the dataset. For example, although 69\% of images in MegaFace had Creative Commons licenses prohibiting commercial use, it is evident from that paper's citations that the dataset was obtained by companies, where there is no way to readily enforce research-only usage. Furthermore, although the MS-Celeb-1M dataset has been taken down by its authors, the data has ``runaway''~\cite{megapixels}---the dataset itself remains available on Academic Torrents, where it continues to be downloaded. And, other researchers have created derivatives of this dataset, which also remain openly accessible online. 

This question of enforcement is important; creators need to consider the potential for misuse of their dataset and what they are in a position to enforce. As illustrated above, it is clear that disclaimers are important to have, but are not enforceable alone. Therefore, in the case of potentially sensitive data, dataset creators should consider reviewing requests on a case-by-case basis, putting forth a good faith effort to reject requests whose intent does not match the purpose of the dataset.

Many datasets do not clearly communicate their limitations nor how they should be used. We have seen some efforts to standardize documentation and increase dataset transparency, notably Datasheets for Datasets~\cite{gebru2018datasheets, holland2018dataset}, as well as an emerging culture of reflective practice to this end within academia and industry. In the past year, we have seen the addition of disclaimers to two existing datasets: the Labeled Faces in the Wild website was updated with a disclaimer about its potential lack of representation, and the VGGFace2 website similarly cautioned that their ``distribution of identities... may not be representative of the global human population'' and that users should ``be careful of unintended societal, gender, racial and other biases''.
While these examples indicate a shift towards greater transparency and acknowledgement of limitations in datasets, they also show that more work remains. Is it enough for dataset owners to qualify the use of datasets, or is it necessary to actively restrict access to and enforce the responsible use of datasets? What role do reflective and interrogatory processes (such as the framework we provide) have in leading to more responsible dataset practices? 

\subsection{Purpose}
\label{sec:purpose}
\emph{Purpose} answers the question, ``Why?'' Philosophers sometimes refer to this component as ``teleology,'' which involves explaining a phenomenon not in terms of what \emph{caused} it but what \emph{motivated} it; teleology covers a range of questions, e.g., what is the dataset for; what are its intended uses; for what purposes is it optimized?

Purpose is an incredibly rich source of ethical concern. A direct challenge to the moral legitimacy of a dataset's own purpose is one such concern. For example, a detractor may assert that the purpose for creating a face dataset being to tell apart gay individuals does not meet a moral threshold~\cite{wang2018deep}. Another way that purpose may stir up ethical concerns is in its relation to other characteristics of a dataset. One cannot overemphasize the potential for ethical discord that may follow. A typical instance is that of a face dataset, not specifically optimized for facial recognition, but used for this purpose. Though it may serve well in some capacities (e.g., face synthesis), it may result in bias, or representational harm to certain minority groups, when used to train a recognition algorithm~\cite{buolamwini2018gender,benjamin2019race,scheuerman2019computers}. Similarly, creators of face datasets considering policies for distributing or providing access to them would want to understand how these policies would apply in relation to certain purposes (e.g., as tools for surveilling subpopulations~\cite{Hassein2017}).

Setting aside cases where a purpose itself is deemed morally reprehensible, purpose (or teleology) stands apart from the other components in that it is frequently \emph{relational} in nature. By this, we mean that ethical issues emerge from mappings---between purpose and properties of creation, distribution, or composition. By implication, fastidious users of our framework will not consider their work complete until they perform a systematic pass of the other three components with a clear sense of a dataset's teleology.

\section{Conclusion}
\label{sec:conclusion}
Our analytic framework extends beyond prior work, bringing into focus different components of dataset creation in which ethical issues may arise. It crystallizes these components, draws explicit threads between them and traditional ethical issues, and demonstrates how those issues manifest. We see our work as part of a broader agenda that strives to reflect critically on the role of academic research in the pursuit and oversight of ethical AI.

The work reported here is the beginning of a longer term effort. In the future, a pressing need is to extract a heuristic from the framework, providing concrete, practical guidance for identifying and mitigating ethical hazards. Furthermore, we will continue refining our framework through collaborations with computer vision researchers developing image datasets. Finally, we will investigate the extensibility of the framework to different modalities and other types of data.

We believe there must be a larger cultural shift within communities of academics and practitioners to acknowledge and address issues in datasets. At the same time, creators should be held accountable for dataset maintenance as limitations are revealed by users and analysts over time. What ultimately is at stake is not only promoting societal values but also maintaining society’s confidence in the work product of the research community in computer vision.

\medskip
\small

\bibliographystyle{unsrt}
\bibliography{04-refs}

\end{document}